\documentclass[twocolumn,showpacs,aps,prd,amsmath,amssymb]{revtex4-2}
\usepackage[dvips]{graphicx}
\usepackage{subfigure}
\usepackage{color}
\usepackage{epsfig}
\usepackage{indentfirst}
\usepackage{multirow}
\usepackage{amsmath}
\usepackage{amssymb}
\usepackage{bm}
\usepackage{fancyhdr}
\usepackage{enumerate}
\usepackage[perpage,symbol]{footmisc}
\usepackage{xspace}
\usepackage{makecell}
\usepackage{float}
\usepackage[bookmarksnumbered,pdfstartview=FitH,colorlinks=true,menucolor=black,linkcolor=blue]{hyperref}

\begin{document}

\title{\boldmath Study of excited $\Xi$ states in $\psi(3686)\rightarrow{}K^{-}\Lambda\overline{\Xi}^{+}+c.c.$}

\author{
M.~Ablikim$^{1}$, M.~N.~Achasov$^{13,b}$, P.~Adlarson$^{73}$, R.~Aliberti$^{34}$, A.~Amoroso$^{72A,72C}$, M.~R.~An$^{38}$, Q.~An$^{69,56}$, Y.~Bai$^{55}$, O.~Bakina$^{35}$, I.~Balossino$^{29A}$
% ...
\\
\vspace{0.2cm}
(BESIII Collaboration)\\
\vspace{0.2cm} {\it
$^{1}$ Institute of High Energy Physics, Beijing 100049, People's Republic of China\\
% ...
$^{13}$ G.I. Budker Institute of Nuclear Physics SB RAS (BINP), Novosibirsk 630090, Russia\\
% ...
$^{29}$ INFN Sezione di Ferrara, (A)INFN Sezione di Ferrara, I-44122, Ferrara, Italy; (B)University of Ferrara,  I-44122, Ferrara, Italy\\
% ...
$^{34}$ Johannes Gutenberg University of Mainz, Johann-Joachim-Becher-Weg 45, D-55099 Mainz, Germany\\
$^{35}$ Joint Institute for Nuclear Research, 141980 Dubna, Moscow region, Russia\\
% ...
$^{38}$ Liaoning Normal University, Dalian 116029, People's Republic of China\\
% ...
$^{55}$ Southeast University, Nanjing 211100, People's Republic of China\\
$^{56}$ State Key Laboratory of Particle Detection and Electronics, Beijing 100049, Hefei 230026, People's Republic of China\\
% ...
$^{69}$ University of Science and Technology of China, Hefei 230026, People's Republic of China\\
% ...
$^{72}$ University of Turin and INFN, (A)University of Turin, I-10125, Turin, Italy; (B)University of Eastern Piedmont, I-15121, Alessandria, Italy; (C)INFN, I-10125, Turin, Italy\\
$^{73}$ Uppsala University, Box 516, SE-75120 Uppsala, Sweden\\
% ...
$^{b}$ Also at the Novosibirsk State University, Novosibirsk, 630090, Russia\\
% ...
}

}
\affiliation{}

\date{\today}

\begin{abstract}

  Based on a sample of $(448.1\pm2.9)\times10^{6}$ $\psi(3686)$ events collected with the
  BES\uppercase\expandafter{\romannumeral3} detector at
  BEPC\uppercase\expandafter{\romannumeral2},
  the decays of
  $\psi(3686)\to{}K^{-}\Lambda\overline{\Xi}^{+} + c.c.$
  with $\overline{\Xi}^+ \to \overline{\Lambda} \pi^+$, $\overline{\Lambda}\to \overline{p} \pi^+$
  are studied.
  We investigate the two excited resonances, $\Xi(1690)^-$ and $\Xi(1820)^-$, which are each observed with large significance ($ \gg 10 \sigma$) in the
  $K^{-}\Lambda$ invariant mass distributions.
  A partial wave analysis is performed, and the spin-parities of $\Xi(1690)^-$ and $\Xi(1820)^-$
  are measured to be $\frac{1}{2}^{-}$ and $\frac{3}{2}^{-}$, respectively. The masses, widths, and
  product branching fractions of $\Xi(1690)^-$ and $\Xi(1820)^-$ are also measured.
\end{abstract}


\maketitle
\section{INTRODUCTION}
In the quark model~\cite{introduction:quarkmodel}, hadrons are viewed as composite objects of constituent spin-$\frac {1}{2}$ quarks bound by the strong interaction. Mesons are made of quark-antiquark ($q \overline q$) pairs and baryons are made of three quarks ($qqq$). Within this simple quark model, the qualitative properties of hadrons and the phenomenology of meson and baryon spectroscopy are well-explained. The accepted full theory of the strong interaction is Quantum Chromodynamics (QCD), a non-Abelian gauge-field theory that describes the interactions of quarks and gluons and has the features of asymptotic freedom and confinement of quarks.
The understanding of the quark-gluon structure of baryons is one of the most important tasks in both particle and nuclear physics.
Since baryons represent the simplest system in which the three colors of QCD neutralize into colorless objects
and the essential non-Abelian character of QCD is manifested, systematic study of baryon spectroscopy can provide critical
insights into the nature of QCD in the confinement domain.

The mass spectra, together with their production and decay rates, provide the main sources of information to study their structure.
Much experimental work has been dedicated to the study of baryon spectroscopy.
However, the available experimental information  for strange baryons remains very incomplete.  In particular,
we are lacking knowledge of the excited baryon states with two strange quarks, {\it i.e.} $\Xi ^*$ hyperons,
due to their small production cross sections and the complicated topology of the final states.
Some phenomenological QCD-inspired models predict more than thirty such kinds of hyperons,
however, only about a dozen total $\Xi$ states have been observed to date.
Among them, only a few are well established with spin-parity determined.
The spin of $\Xi(1820)$ is determined to be $\frac{3}{2}$  ~\cite{introduction:teodo78} and the corresponding parity is   measured to be negative ~\cite{introduction:biagi87c}.
Some evidence that the $\Xi(1690)$ has $J^P=\frac{1}{2}^-$ was found in a study of $\Lambda^+_c\rightarrow\Xi^- \pi^+ K^+$ ~\cite{introduction:aubert08}.

In recent years, charmonium data samples with unprecedented statistics were accumulated
by the Beijing Spectrometer (BESIII~\cite{detector:bes3details})
at the Beijing Electron-Positron Collider (BEPCII~\cite{detector:bepc2}),
and these provide great opportunities for investigating the light baryons produced in charmonium decays.
In a previous analysis using a sample of $106 \times 10^{6}$ $\psi(3686)$ events collected with BESIII,
two hyperons, $\Xi(1690)^-$ and $\Xi(1820)^-$, were observed
in the $K^{-}\Lambda$ invariant mass distribution~\cite{intoduction:preresults}.
Now, with four times more $\psi(3686)$ events collected at BESIII,
we conduct a more extensive study of the decays $\psi(3686)\to{}K^{-}\Lambda\overline{\Xi}^{+} + c.c.$.
In particular, we perform a partial wave analysis (PWA) to study the properties of intermediate state $\Xi^{*}$ hyperons.

In this paper, we report a PWA analysis of $\psi(3686)\to{}K^{-}\Lambda\overline{\Xi}^{+} + c.c.$,
with a sample of $(448.1\pm2.9)\times10^{6}$ $\psi(3686)$ events collected at BESIII.
In the following, the charge conjugate channel is always implied.

\section{BESIII DETECTOR}
The BESIII detector records symmetric $e^+e^-$ collisions
provided by the BEPCII storage ring,
which operates in the center-of-mass energy range from 2.0 to 4.95~GeV.
BESIII has collected large data samples in this energy region~\cite{detector:whitepaper}.
The cylindrical core of the BESIII detector covers 93\% of the full solid angle and
consists of a helium-based multilayer drift chamber~(MDC),
a plastic scintillator time-of-flight system~(TOF),
and a CsI(Tl) electromagnetic calorimeter~(EMC),
which are all enclosed in a superconducting solenoidal magnet
providing a 1.0~T (0.9~T in 2012) magnetic field.
The solenoid is supported by an octagonal flux-return yoke with resistive plate counter muon
identification modules interleaved with steel.
The charged-particle momentum resolution at $1~{\rm GeV}/c$ is
$0.5\%$, and the dE/dx resolution is $6\%$ for electrons
from Bhabha scattering. The EMC measures photon energies with a
resolution of $2.5\%$ ($5\%$) at $1$~GeV in the barrel (end cap)
region.
The time resolution of the
TOF in the barrel region is 68 ps
while that in the endcap region is 110 ps.

\section{DATA SETS AND MONTE CARLO SAMPLES}
This study uses $(448.1 \pm 2.9) \times 10^6$
$\psi(3686)$ events collected by the BESIII detector at BEPCII in 2009 ($(107.0 \pm 0.8)\times 10^6$ events) and 2012 ($(341.1 \pm 2.1)\times 10^6$ events, taken with 0.9 T magnetic field) ~\cite{detector:psip}.

A GEANT4-based~\cite{detector:geant4} Monte Carlo (MC) simulation software BOOST~\cite{detector:bes3sim},
which includes a geometric and material description~\cite{detvis} of the BESIII detector, detector response and digitization models
as well as tracking of the detector running conditions and performance, is used to generate MC simulated data samples.
An exclusive MC sample for the process $\psi(3686)\rightarrow{}K^{-}\Lambda\overline{\Xi}^{+}$,
is generated to optimize the selection criteria and estimate the corresponding selection efficiency.
The production of the $\psi(3686)$ is simulated by the generator KKMC~\cite{detector:kkmc}, and the subsequent decays are generated with BesEvtGen~\cite{detector:evtgen,detector:evtgen2}.
An inclusive MC sample, consisting of $448\times10^{6}$ $\psi(3686)$ events, is used to study potential backgrounds.
The known decay modes of the $\psi(3686)$ are generated by BesEvtGen
with branching fractions set to world average values~\cite{introduction:2018pdg},
and the remaining unknown decay modes are modeled by LUNDCHARM~\cite{detector:lundcharm}~\cite{LUND2}.

\section{EVENT SELECTION}
Considering the full decay chain of $\psi(3686)\rightarrow{}K^{-}\Lambda\overline{\Xi}^{+}$ as reconstructed from the decays $\Lambda\rightarrow{} p \pi^-$,
$\overline{\Xi}^{+}\rightarrow\overline{\Lambda}\pi^{+}$ and
$\overline{\Lambda}\rightarrow\overline{p}\pi^{+}$,
there are six charged tracks with low momentum in the final state, and the detection efficiency is very low.
Therefore a partial reconstruction method is adopted to obtain higher statistics
by not requiring the prompt $\Lambda$ from the $\psi(3686)$ decay.
Following the $K^-$ and $\overline{\Xi}^{+}$ reconstruction, the $\Lambda$ four-momentum is calculated from the recoil of the $K^{-}\overline{\Xi}^{+}$ system.

With the partial reconstruction method, at least four charged tracks are required.  The polar angles $\theta$, defined with respect to the axis of the MDC, of all charged tracks is required to satisfy $|\cos\theta|<0.93$.
For the kaon, the point of closest approach to the beam line is required to be within $\pm10$~cm in the beam direction and 2~cm in the plane perpendicular to the beam.
Since the $\overline{\Xi}^{+}$ particle has a displaced decay vertex,
looser requirements are imposed on the charged tracks from the $\overline{\Xi}^{+}$ decay: the point of closest approach to the beam line of is only required to be within $\pm15$~cm in the beam direction and 10~cm in the plane perpendicular to the beam.

Particle identification~(PID) for charged tracks combines measurements of the energy loss in the MDC~(d$E$/d$x$) and the flight time in the TOF to form likelihoods $\mathcal{L}(h)~(h=p,K,\pi)$ for each hadron $h$ hypothesis.
Tracks are identified as protons when the proton hypothesis has the highest likelihood ($\mathcal{L}(p)>\mathcal{L}(K)$ and $\mathcal{L}(p)>\mathcal{L}(\pi)$), while charged kaons and pions are identified by comparing the likelihoods for the kaon and pion hypotheses, $\mathcal{L}(K)>\mathcal{L}(\pi)$ and $\mathcal{L}(\pi)>\mathcal{L}(K)$, respectively.
In this analysis, two negatively charged tracks are required to be identified as $K^-$ and $\overline{p}$,
and two positively charged tracks as pions.

\begin{figure*}[htbp]
\centering
\includegraphics[width=0.39\textwidth]{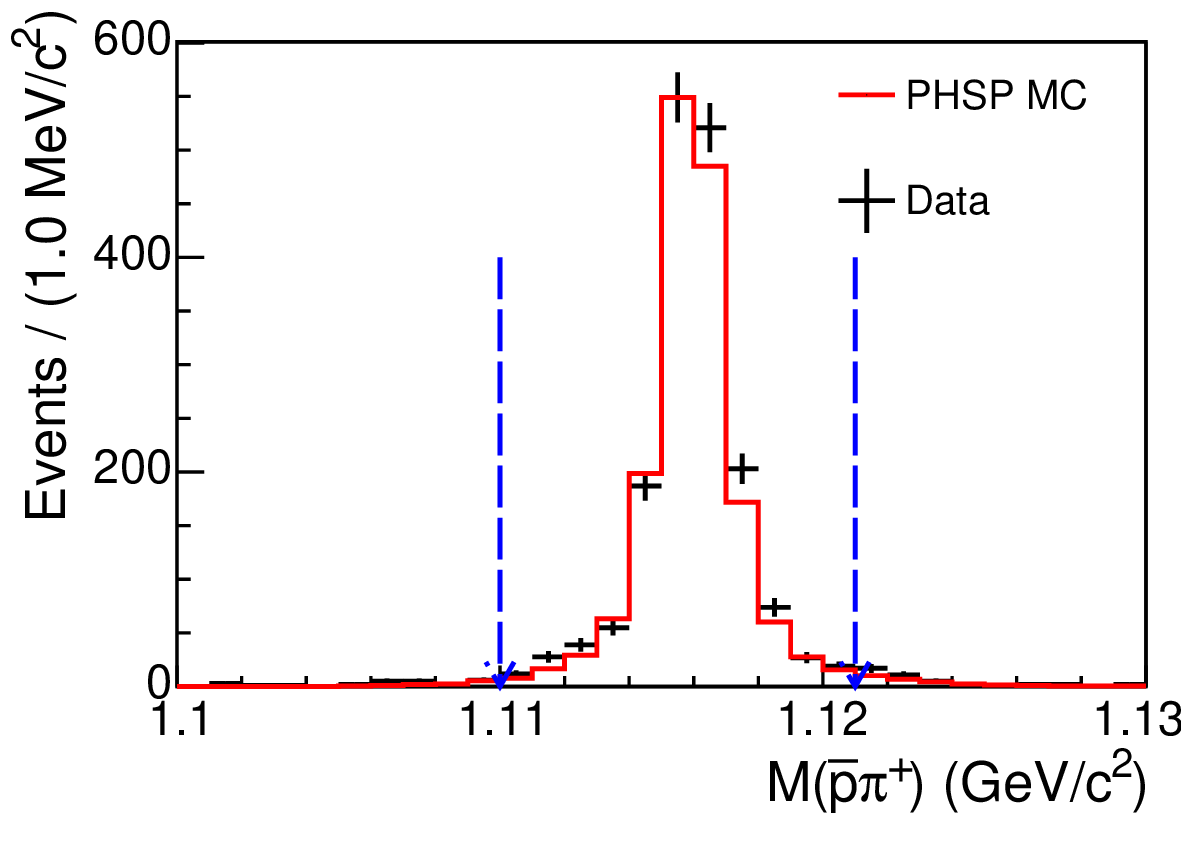}
\put(-155,110){(a)}
\includegraphics[width=0.39\textwidth]{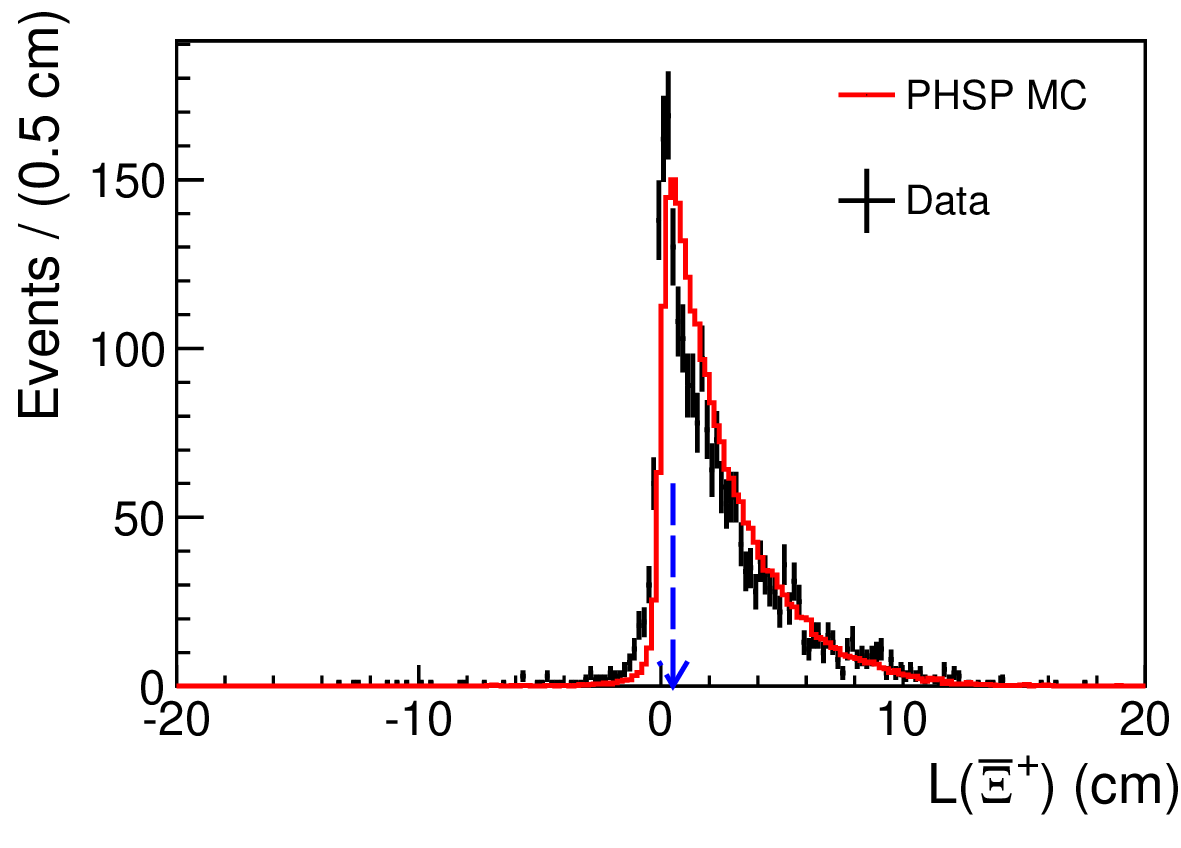}
\put(-155,110){(b)}\\
\includegraphics[width=0.39\textwidth]{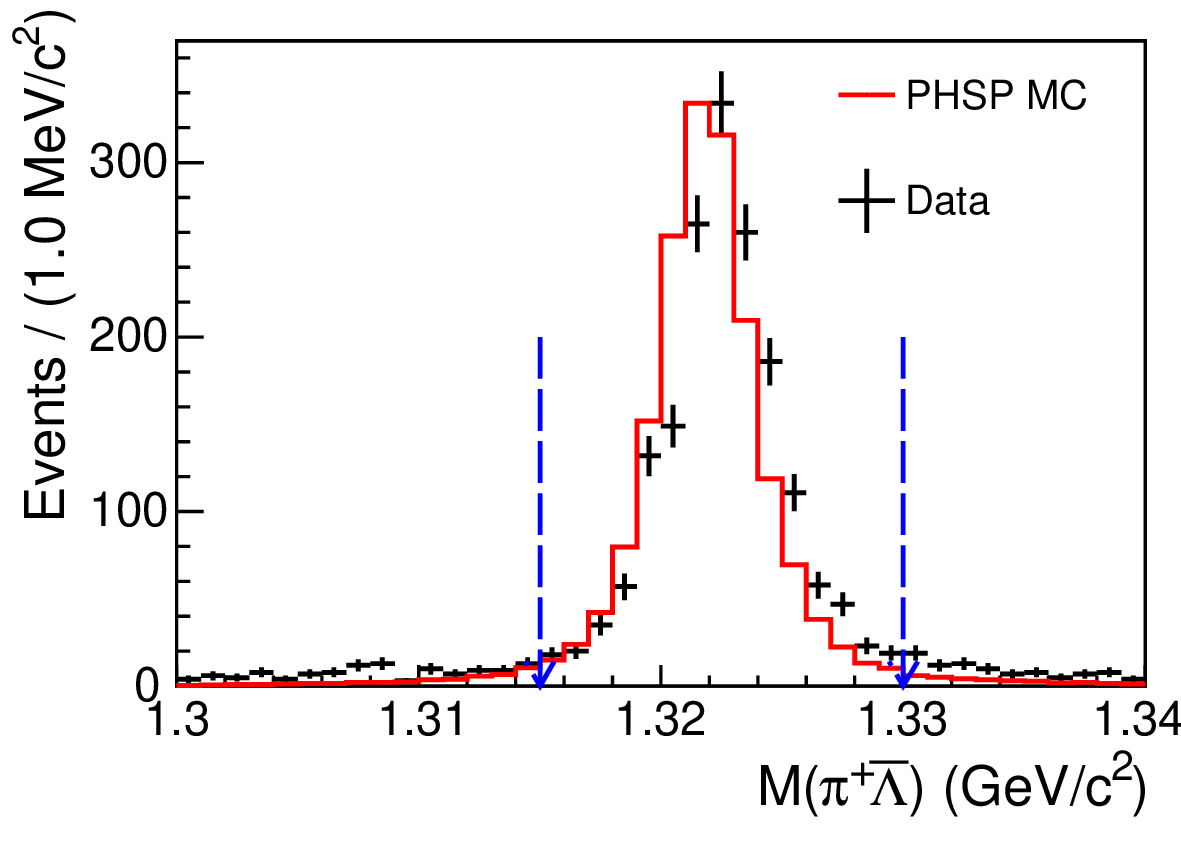}
\put(-155,110){(c)}
\includegraphics[width=0.39\textwidth]{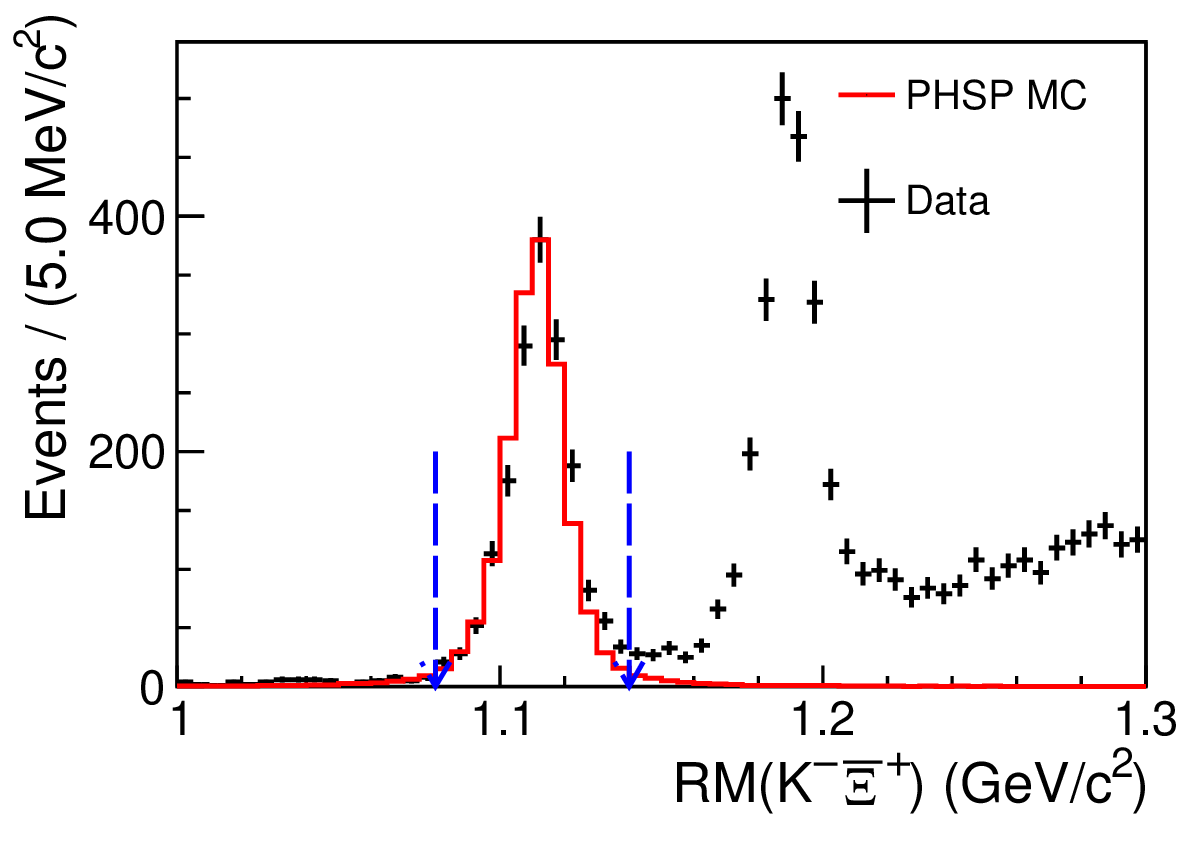}
\put(-155,110){(d)}
\caption{
	 Distribution of $M(\overline{p} \pi^+)$ of $\overline{\Lambda}$ candidates (a),
	 decay length of $\overline{\Xi}^+$ candidates (b), and
	 $M(\overline{\Lambda}\pi^+)$ of $\overline{\Xi}^+$ candidates (c).
    Plot (d) shows the distribution of RM$(K^-\overline{\Xi}^+)$: the left peak is $\Lambda$, while the right one is $\Sigma^0$.
	The crosses represent data and the histograms represent phase space (PHSP) MC. The dashed arrows show the cut values; for each plot, cuts on the other three quantities are applied.  }
\label{fig:evtslt_cuts}
\end{figure*}

The candidate $\overline{\Xi}^{+}$ baryon is reconstructed in two steps.
A $\overline{p}\pi^{+}$ pair sharing a common vertex is selected to reconstruct the $\overline{\Lambda}$ candidate
via a secondary vertex fit.
The $\overline{\Xi}^{+}$ is then reconstructed with the $\overline{\Lambda}$ candidate
and the other $\pi^{+}$ by applying another secondary vertex fit.
For events with more than one $\overline{\Xi}^{+}$ candidate,
the $\overline{p}\pi^{+}$ combination with the minimum $|M(\overline{p}\pi^{+})-M(\overline{\Lambda})|$ is selected.
Here, $M(\overline{p}\pi^{+})$ is the invariant mass of the $\overline{p}\pi^+$ combination,
see Fig.~\ref{fig:evtslt_cuts} (a), and
$M(\overline{\Lambda})$ is the known mass of $\Lambda$ taken from the Particle Data Group (PDG)~\cite{introduction:2018pdg}.

A mass window of $1.110\;\text{GeV}/c^{2}<M(\overline{p}\pi^{+})<1.121\;\text{GeV}/c^{2}$ is imposed
to select $\overline{\Lambda}$ candidates as shown
in Fig.~\ref{fig:evtslt_cuts} (a) by the blue dashed arrows.
The distribution of the $\overline{\Xi}^{+}$ decay length, $L(\overline{\Xi}^+)$, is shown in Fig.~\ref{fig:evtslt_cuts} (b).
The distribution of the $\overline{\Lambda}\pi^{+}$ invariant mass, $M(\overline{\Lambda}\pi^{+})$, is shown in Fig.~\ref{fig:evtslt_cuts} (c), after a requirement of $L(\overline{\Xi}^{+})>0.5~\text{cm}$ is applied.
The invariant mass $M(\overline{\Lambda}\pi^{+})$ is required to satisfy
$1.315\;\text{GeV}/c^{2}<M(\overline{\Lambda}\pi^{+})<1.330\;\text{GeV}/c^{2}$.
The distribution of the mass recoiling against $K^{-}$ and
the reconstructed $\overline{\Xi}^{+}$, RM($K^-\overline{\Xi}^+$), is shown in Fig.~\ref{fig:evtslt_cuts} (d).
One can see a clear $\Lambda$ baryon signal around $1.115$~GeV/$c^2$ and the main background from $\Sigma^0$ decays around $1.193$~GeV/$c^2$.
The requirement $1.080\;\text{GeV}/c^{2}< \text{RM}(K^{-}\overline{\Xi}^{+})<1.140\;\text{GeV}/c^{2}$ is imposed
to select prompt $\Lambda$ candidates.  In total, 1714 events are selected.

\begin{figure*}[htbp]
  \centering
  	\includegraphics[height=50mm,width=0.39\textwidth]{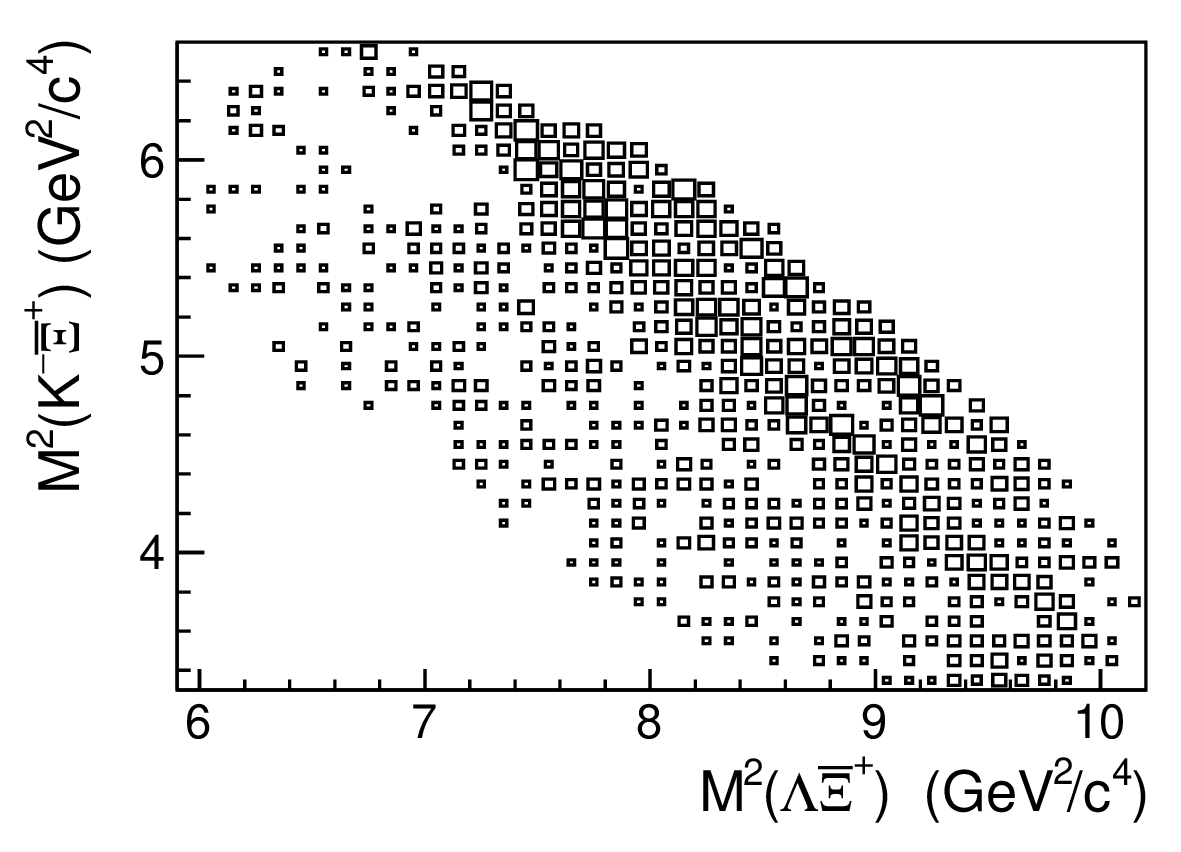}
  \put(-30,120){(a)}
  	\includegraphics[width=0.39\textwidth]{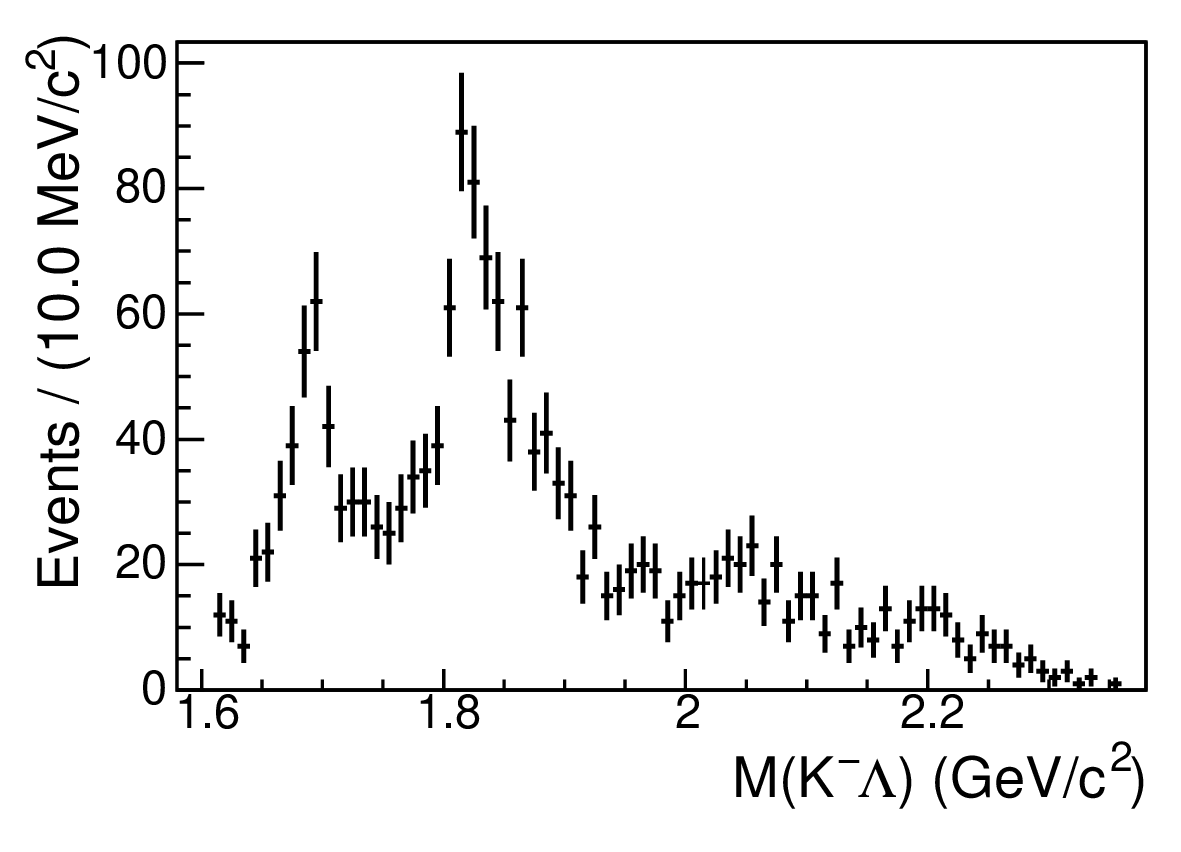}
   \put(-30,120){(b)}\\
  	\includegraphics[width=0.39\textwidth]{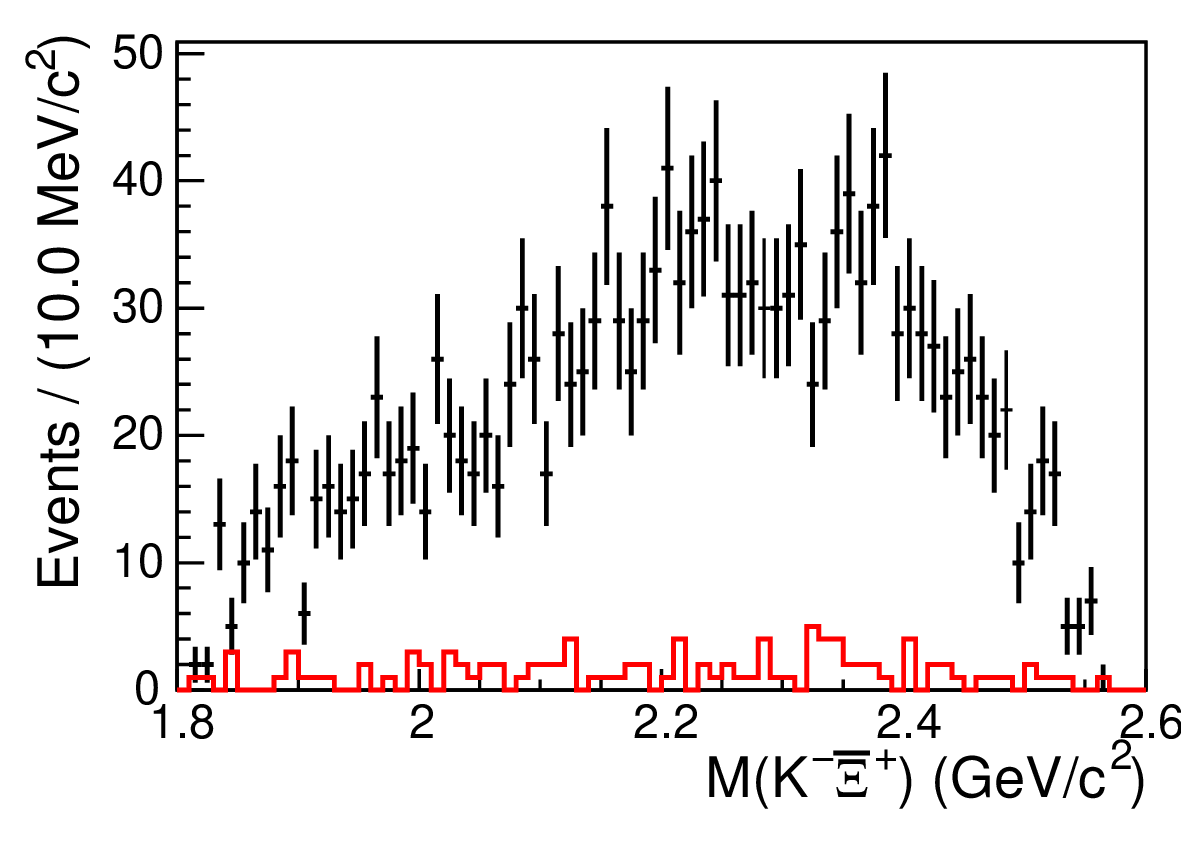}
  \put(-30,120){(c)}
  	\includegraphics[width=0.39\textwidth]{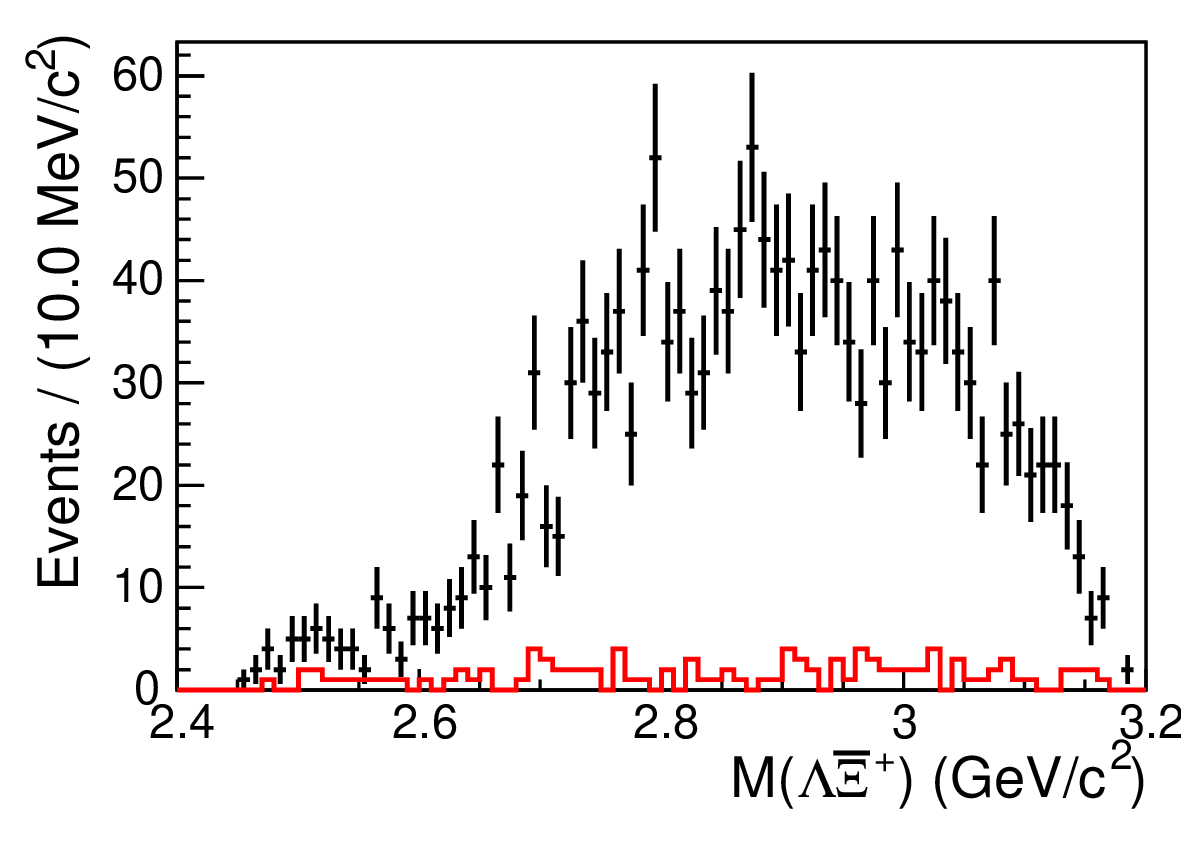}
  \put(-30,120){(d)}
  \caption{
  (a) Dalitz plot of $M^{2}(\Lambda\overline{\Xi}^{+})\; vs \;M^{2}(K^{-}\overline{\Xi}^{+})$ and the distributions of (b) $M(K^{-}\Lambda)$, (c) $M(K^{-}\overline{\Xi}^{+})$ and (d) $M(\Lambda\overline{\Xi}^{+})$ after the final selection. The crosses represent the data and the histograms represent the background events estimated from the $\Xi$ sidebands.
  }
  \label{fig:evtslt_spectrums}
\end{figure*}

After the above selection,
a one-constraint kinematic fit that constrains the mass of the missing $\Lambda$ to the PDG value,
is performed to improve the resolution, and no event is rejected.
The distributions of $M(K^-\overline{\Xi}^+)$, $M(\Lambda\overline{\Xi}^+)$, and $M(K^-\Lambda)$
are shown in Fig.~\ref{fig:evtslt_spectrums} (b), (c) and (d), respectively.
Two diagonal bands on the Dalitz plot as shown in Fig.~\ref{fig:evtslt_spectrums} (a) correspond to
the two structures near 1.7~GeV/$c^{2}$ and 1.8~GeV/$c^{2}$ in the $M(K^{-}\Lambda)$ mass spectrum.

To investigate possible background events,
the same analysis is performed to the $\psi(3686)$ inclusive MC sample,
and nineteen background events are found. A detailed event type analysis with a generic tool, TopoAna ~\cite{incbg}, shows that the
 main background is from $\psi(3686)\rightarrow{}\gamma\chi_{c1,c2},
\chi_{c1,c2}\rightarrow{}n\overline{K}^{*}\overline{\Lambda}+c.c.$ decays
plus some other decays with $K^-\pi^+\pi^-\overline p$ in the final state.
A sideband method is used in data to estimate the background contribution.
The sideband regions of $\overline \Xi^+$  are defined to be
 $(1.3025,1.3100)
$ and $(1.3350,1.3425)$ GeV/$c^{2}$.
A total of  104 background events, corresponding to a background level of 6\%,  are obtained.  No peaking background is observed in the recoil mass for the $\overline \Xi^+$ sideband events.  
The lower background level in the MC is attributed to the lack of simulation of the decays of higher-mass excited states and to the incomplete description of the decays of the excited states that are simulated.
Additionally, the continuum data taken at a center-of-mass energy of 3.65 GeV with an integrated luminosity of 42.6~pb$^{-1}$ is used to estimate the background from Quantum Electrodynamics processes, and no background event remains.

\section{PARTIAL WAVE ANALYSIS}
The two-body decay amplitudes in the sequential decay process
$\psi(3686)\rightarrow\overline{\Xi}^{+}\Xi^{*-}_{X}, \Xi^{*-}_{X}\to{}K^{-}\Lambda$
are constructed using the relativistic covariant tensor amplitude formalism~\cite{pwa:amplitude},
and the maximum likelihood method is used in the PWA, with the FDC~\cite{pwa:fdc} package. Here, $\Xi^{*-}_{X}$ denotes an intermediate state such as $\Xi(1620)^-$, $\Xi(1690)^-$, $\Xi(1830)^-$, etc.

\subsection{Introduction to PWA}
The amplitude $A_{j}$ for the $j$-th possible partial wave in
$\psi(3686)\rightarrow\overline{\Xi}^{+}\Xi^{*-}_{X}, \Xi^{*-}_{X}\to{}K^{-}\Lambda$
is described as
\begin{equation}
  A_{j}=A^{j}_{{\rm prod},X}(BW)_{X}A_{{\rm decay},X},
\end{equation}
where $A^{j}_{{\rm prod},X}$ is the amplitude describing the production
of the intermediate resonance $\Xi^{*-}_{X}$,
$BW_{X}$ is the Breit-Wigner propagator of $\Xi^{*-}_{X}$,
and $A_{{\rm decay},X}$ is the decay amplitude of $\Xi^{*-}_{X}$.

  The total differential cross section $d\sigma/d\Phi$ is
\begin{equation}
  \frac{d\sigma}{d\Phi}= \left| \sum\limits_{j}c_{j}A_{j} \right| ^{2},
\end{equation}
where $\sigma$ is the total cross section, $\Phi$ is the phase space, and $c_{j}$ is a complex free parameter to be determined in the fit for each partial wave $A_{j}$.

The probability to observe the event characterized by the variable $\xi$ is
\begin{equation}
  P(\xi)=\frac{\omega(\xi) \, \epsilon(\xi)}{\int d\xi \, \omega(\xi) \, \epsilon(\xi)},
\end{equation}
where $\xi$ are the four-momenta of the $K^{-}$, $\Lambda$, and $\overline{\Xi}^{+}$,
$\omega(\xi) \equiv d\sigma/d\Phi$ is the probability density for a single event to
populate the PHSP at $\xi$, and
$\epsilon(\xi)$ is the detection efficiency to detect one event with $\xi$.
The normalization integral $\int d\xi \, \omega(\xi) \, \epsilon(\xi)$ is calculated using the exclusive signal MC sample.

The likelihood for observing $N$ events in the data sample is
\begin{align}
  \mathcal{L}=P(\xi_{1},\xi_{2},...,\xi_{N})&=\prod\limits_{i=1}^{N}P(\xi_{i}) \notag\\
  &=
  \prod\limits_{i=1}^{N}\frac{\omega(\xi_{i}) \, \epsilon(\xi_{i})}{\int d\xi \, \omega(\xi) \, \epsilon(\xi)}.
\end{align}

Rather than maximizing the likelihood function $\mathcal{L}$, the
quantity $-\ln\mathcal{L}$ is minimized to obtain best values of the parameters $c_{j}$
and the masses and widths of the resonances
\begin{equation}
  -\ln\mathcal{L}=-\sum\limits_{i=1}^{N}\ln\left(\frac{\omega(\xi_{i})}
  {\int d\xi \, \omega(\xi) \, \epsilon(\xi)}\right)
  -\sum\limits_{i=1}^{N}\ln\epsilon(\xi_{i}),
\end{equation}

For a given data set, the second term is a constant and has no impact on the
determination of the parameters of the amplitudes or on the change of $-\ln\mathcal{L}$. So, in the fit, the $-\ln\mathcal{L}$ is defined as
\begin{equation}
  -\ln\mathcal{L}=-\sum\limits_{i=1}^{N}\ln\left(\frac{\omega(\xi_{i})}
  {\int d\xi \, \omega(\xi) \, \epsilon(\xi)}\right).
\end{equation}

The final log-likelihood value minimized for the data, $S$, is the sum of the log-likelihood values of the events in the $\Xi$ signal region and
background events in the $\Xi$ sideband region with negative weights.
\begin{equation}
  S=-\ln\mathcal{L}_{\rm{data}} + \ln\mathcal{L}_{\rm{bg}}.
  \label{eq:likelihood}
\end{equation}

The free parameters are optimized by FUMILI~\cite{pwa:fumili}.
In the minimization procedure, a change in the log-likelihood of 0.5 represents one
standard deviation for each parameter.

In this analysis, the Breit-Wigner resonance shape used for the $\Xi^{*}$ is
\begin{equation}
  BW(s)=\frac{1}{M_{\Xi^{*}}^{2}-s_{K\Lambda}-iM_{\Xi^{*}}\Gamma_{\Xi^{*}}},
\end{equation}
where $s_{K\Lambda}$ is the invariant mass squared of the decay products of the $\Xi^*$.

Since nucleons have structure, form factors modifying the Breit-Wigner shape are needed to describe them.
Different form factors have been discussed in Refs.~\cite{pwa:formfactor1, pwa:formfactor2},
and the following ones are used in the fit:
\begin{align}
  J&=\frac{1}{2}: &\quad F_{N}(s_{K\Lambda})&=\frac{\lambda^{4}_{1}}
  {\lambda^{4}_{1}+(s_{K\Lambda}-M_{\Xi^{*}}^{2})^{2}},
  \label{eq:ff1} \\
  J&=\frac{3}{2}, \frac{5}{2}: &\quad
  F_{N}(s_{K\Lambda})&=e^{\frac{-|s_{K\Lambda}-M_{\Xi^{*}}^{2}|}
  {\lambda^{2}_{2}}},
  \label{eq:ff2}
\end{align}
where $J$ is the spin, and the cutoff parameters $\lambda_{1}$ and $\lambda_{2}$ are set to be 2.0 GeV.

\subsection{PWA results}
According to the $M(K^{-}\overline{\Xi}^{+})$, $M(\Lambda\overline{\Xi}^{+})$,
and $M(K^{-}\Lambda)$ spectra, as shown in Fig.~\ref{fig:evtslt_spectrums},
no significant enhancement is observed except the two structures
near 1.7~GeV/$c^{2}$ and 1.8~GeV/$c^{2}$ in the $M(K^{-}\Lambda)$ spectrum.
In the PWA,  there are seven PDG-listed candidate $\Xi$ hyperons,
$\Xi(1620)^-$, $\Xi(1690)^-$, $\Xi(1820)^-$, $\Xi(1950)^-$, $\Xi(2030)^-$, $\Xi(2120)^-$ and $\Xi(2250)^-$.
A coherent non-resonant contribution is also considered, denoted as non-res, which is described as a wide intermediate state  with certain spin-parity.

\subsubsection{Nominal fit}
In the first step of partial wave analysis, all possible sets of amplitudes corresponding to the seven PDG-listed candidate $\Xi$ hyperons  are evaluated. The masses and widths of resonances are fixed to the PDG. The significance for a resonance is calculated based on the improvement in PWA quality, $\Delta S$, with the change in degrees of freedom considered.
Only the $\Xi(1690)^-$ and $\Xi(1820)^-$ have significances greater than 5$\sigma$.
The other five $\Xi$ resonances tested ($\Xi(1620)^-$, $\Xi(1950)^-$, $\Xi(2030)^-$, $\Xi(2120)^-$ and $\Xi(2250)^-$), each with significance less than 5$\sigma$, are excluded from the nominal fit.  Each was tried with a variety of $J^P$ values: $1/2^\pm, 3/2^\pm, 5/2^\pm, 7/2^\pm$.
Their impact will be considered as a systematic uncertainty. The $J^P$ of  $\Xi(1690)$ and $\Xi(1820)$ are favored to be $\frac{1}{2}^-$  and $\frac{3}{2}^-$,  respectively.

In the next step, the masses and widths of $\Xi(1690)$ and $\Xi(1820)$ are further optimized. The obtained results are shown in Table~\ref{table:pwa_opt}.
The masses and widths of $\Xi(1690)^-$ and $\Xi(1820)^-$ are consistent with the PDG values within $2.6 \sigma$.

\begin{figure*}[!htbp]
  \centering
  \includegraphics[width=0.90\textwidth]{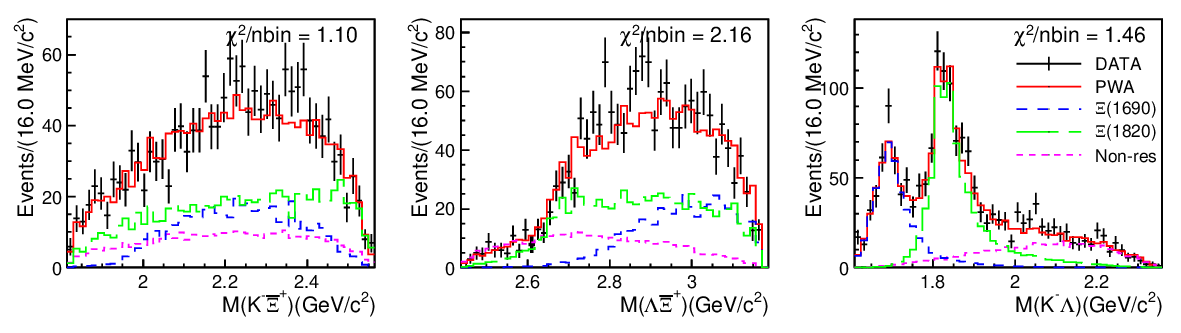}
  \put(-420,110){(a)}
   \put(-270,110){(b)}
    \put(-115,110){(c)}
  \caption{Distributions of (a) $M({K^{-}\overline{\Xi}^{+}})$,
  (b) $M({\Lambda\overline{\Xi}^{+}})$ and (c) $M({K^{-}\Lambda})$. The crosses represent data and red solid histograms represent the projection of the PWA result. The different color histograms represent the intensity of each component in the nominal fit.
  Here, $\chi^2/{\rm nbin}$ demonstrates the goodness of fit in each figure,
  where nbin is the number of bins in each figure and
  $\chi^{2}$ is defined as $\chi^{2}=\sum_{i=1}^{\rm nbin} \, (n_i - v_i)^2 / v_{i}$,
  where $n_{i}$ and $v_{i}$ are the numbers of events for the data and the fit projections of the nominal fit in the $i^{th}$ interval of each figure, respectively.}
  \label{fig:results_mass}
\end{figure*}

The projections on
the $M(K^{-}\overline{\Xi}^{+})$,
$M(\Lambda\overline{\Xi}^{+})$, $M(K^{-}\Lambda)$ spectra after PWA are shown in Fig.~\ref{fig:results_mass}.
They agree with those of the data.
We observe $464\pm43$ $\Xi(1690)^-$ events
with a mass $M=1685^{+3}_{-3}$ MeV/$c^{2}$ and a width $\Gamma=81^{+10}_{-9}$ MeV,
and $776\pm42$ $\Xi(1820)^-$ events
with a mass $M=1821^{+2}_{-3}$ MeV/$c^{2}$ and a width $\Gamma=73^{+6}_{-5}$ MeV.
Here, the uncertainties are statistical only.
The statistical significances of both structures are greater than $10\sigma$.
These significances as well as the fit fractions are given in Table~\ref{table:pwa_opt}.

\begin{table}[!htbp]
  \caption{The optimized mass and width for each resonance, along with the fit fraction (FF) for each. Here, the uncertainties are statistical only. }
  \label{table:pwa_opt}
  \centering
  \begin{tabular}{cccccc}
    \hline
    \hline
    Resonance & $J^{P}$ & M (MeV/$c^2$) & $\Gamma$ (MeV) & $\sigma$ & FF\\
    \hline
    $\Xi(1690)^-$ & $1/2^{-}$ & $1685^{+3}_{-3}$ & $81^{+10}_{-9}$ & 10.8 & 29.0\\
    $\Xi(1820)^-$ & $3/2^{-}$ & $1821^{+2}_{-3}$ & $73^{ +6}_{-5}$ & 18.3 & 48.0\\
    Non-res          & $1/2^{+}$ &    -             &    -         & \(>\)30 & 23.0\\
    \hline
    \hline
  \end{tabular}
\end{table}

\subsubsection{Check of the nominal fit}
Different $J^{P}$ assignments for the
nominal fit
have been tested as shown in Table~\ref{table:pwa_checkSP}.
The likelihood values become worse with respect to that of the
nominal fit.
\begin{table}[htbp]
  \caption{The checks of different $J^{P}$ assignments. $\Delta{}S$ is the change of $S$ compared to the nominal fit.}
  \label{table:pwa_checkSP}
  \centering
  \begin{tabular}{cccc}
    \hline
    \hline
    \cline{1-3} $J^P$ & Non-res$\to{}K^{-}\Lambda$ &~ $\Xi(1690)^-$ ~& $\Xi(1820)^-$ \\
    \hline
    $1/2^{-}$ & 53.3 &  -    & 11.2 \\
    $1/2^{+}$ &  -   & 29.2  & 12.6 \\
    $3/2^{-}$ & 44.9 & 110.6 & -    \\
    $3/2^{+}$ & 7.7  & 33.6  & 13.9 \\
    \hline
    \hline
  \end{tabular}
\end{table}

The other possible non-resonant contributions in the $\Lambda\overline{\Xi}^{+}$ and $K^{-}\overline{\Xi}^{+}$
systems are investigated by replacing $1/2^{-}$ non-res$\rightarrow{}K^{-}\Lambda$ in the
nominal fit.
As shown in Table~\ref{table:pwa_checkPHSP}, the likelihood values also become worse.

\begin{table}[htbp]
  \caption{The checks of other possible non-resonant contributions.}
  \label{table:pwa_checkPHSP}
  \centering
  \begin{tabular}{ccc}
    \hline
    \hline
    Non-res process & $J^{P}$ & $\Delta{}S$ \\
    \hline
    Non-res$\to{}K^{-}\Lambda$ & $1/2^{-}$ & 53.3 \\
    \hline
    \multirow{3}{*}{Non-res$\to{}\Lambda\overline{\Xi}^{+}$} & $0^{-}$ & 43.9 \\
     & $1^{-}$ & 14.8 \\
     & $1^{+}$ & 22.3 \\
    \hline
    \multirow{2}{*}{Non-res$\to{}K^{-}\overline{\Xi}^{+}$} & $1/2^{-}$ & 8.8 \\
     & $1/2^{+}$ & 33.9 \\
    \hline
    \hline
  \end{tabular}
\end{table}

\subsubsection{Branching fractions}
To determine the detection efficiencies of $\psi(3686)\rightarrow\Xi(1690)^-\overline{\Xi}^{+}$ and
$\psi(3686)\rightarrow\Xi(1820)^-\overline{\Xi}^{+}$, signal MC events are generated using the PWA amplitude for each process.
The product branching fraction of
$\psi(3686)\rightarrow\overline{\Xi}^{+}\Xi^{*-}(\Xi^{*-}\rightarrow{}K^{-}\Lambda)+c.c.$
is calculated with
\begin{equation}
\begin{aligned}
&  \mathcal{B}(\psi(3686)\rightarrow\overline{\Xi}^{+}\Xi^{*-}+c.c.)\cdot
  \mathcal{B}(\Xi^{*-}\rightarrow{}K^{-}\Lambda) \\
&  =\frac{N_{\Xi^{*}}}{N_{\psi(3686)} \cdot
  \mathcal{B}(\overline{\Xi}^{+} \to \overline{\Lambda}\pi^{+}) \cdot
  \mathcal{B}(\overline{\Lambda} \to \overline{p}\pi^{+}) \cdot \epsilon_{\Xi^{*}}}, \\
\end{aligned}
\end{equation}
where $N_{\Xi^{*}}$ is the number of $\psi(3686)\rightarrow\Xi^{*-}\overline{\Xi}^{+}+c.c.$ events,
 $N_{\psi(3686)}=(448.1\pm2.9)\times10^{6}$ is the total number of $\psi(3686)$
events~\cite{error:psipnum}, and
$\mathcal{B}(\overline{\Lambda} \to \overline{p}\pi^{+})$ and
$\mathcal{B}(\overline{\Xi}^{+} \to \overline{\Lambda}\pi^{+})$ are the corresponding decay branching
fractions~\cite{introduction:2018pdg}.
The detection efficiency is $\epsilon_{\Xi^{*}} = 16.0\%$ for $\Xi(1690)^-$, and the corresponding product branching fraction is $(1.06\pm0.10)\times10^{-5}$. Similarly, the product branching fraction for
$\Xi(1820)^-$ is $(1.78\pm0.10)\times10^{-5}$ with a detection efficiency $\epsilon_{\Xi^{*}}=14.6\%$.

\begin{figure}[htbp]
  \centering
  \includegraphics[width=0.79\columnwidth]{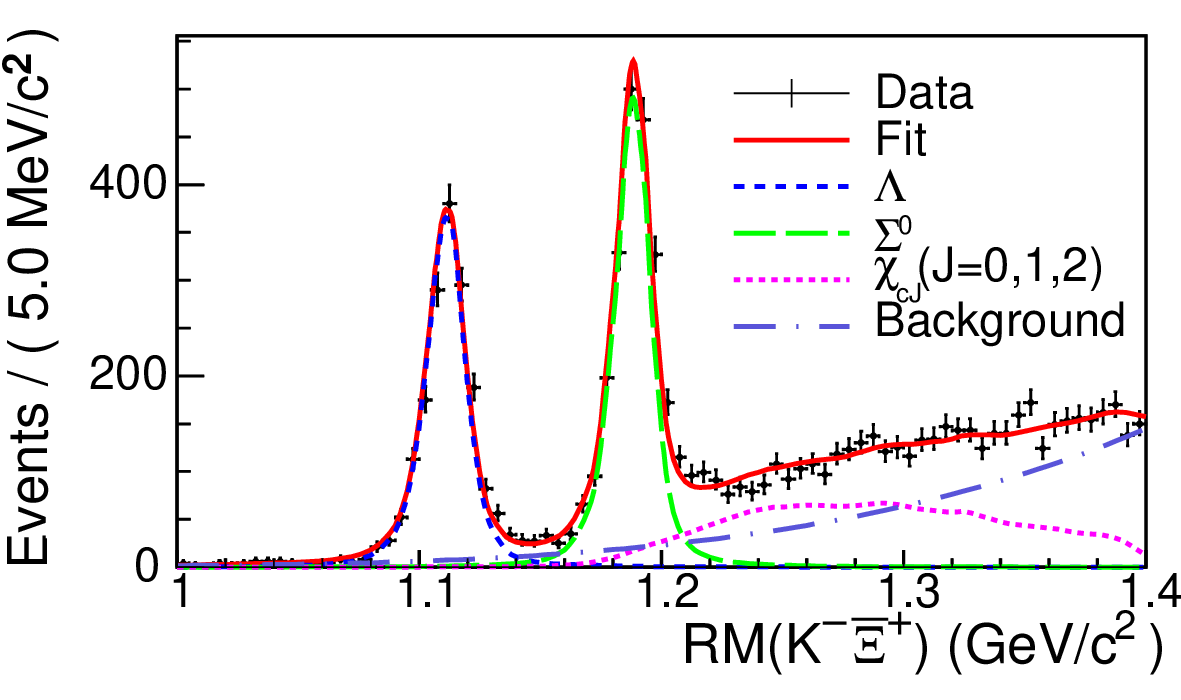}
  \caption{Fit to the RM($K^- \overline{\Xi}^+)$ distribution. The crosses are  data and the red curve denotes the best fit. The other curves show the different fit components listed in the legend.  }
  \label{missingfit}
\end{figure}

Fitting the RM($K^- \overline{\Xi}^+)$ distribution, shown in Fig.~\ref{missingfit}, yields the number of signal events of $\psi(3686)\to K^{-}\Lambda\overline{\Xi}^{+}$ as $N_{\rm sig}= 1572 \pm 45$.
The signal is modeled by a signal MC shape of $\Lambda$ convolved with a Gaussian function while the
background components from $\Sigma^0$ and $\chi_{cJ}$ are described by the MC shapes, and other background channels are described with a third order polynomial function.
The branching fraction of this decay is determined to be
\begin{equation}
\begin{aligned}
&\mathcal{B}(\psi(3686)\to{}K^{-}\Lambda\overline{\Xi}^{+} + c.c.)\\
&  =\frac{N_{\rm sig}}{N_{\psi(3686)} \cdot
  \mathcal{B}(\overline{\Xi}^{+} \to \overline{\Lambda}\pi^{+}) \cdot
  \mathcal{B}(\overline{\Lambda} \to \overline{p}\pi^{+}) \cdot \epsilon} \\
&  = (3.60\pm0.10)\times10^{-5}.
\end{aligned}
\end{equation}
Here, $\epsilon=15.3\%$ is the detection efficiency for the final state.
It is studied with the exclusive signal MC events which are generated using the
PWA results obtained in this analysis.  The uncertainty is statistical only.

\section{SYSTEMATIC UNCERTAINTIES}
In this analysis, the sources of the systematic uncertainty are classified into two categories:
the uncertainty from event selection and the uncertainty from the PWA procedure.
The former affects the measurement of branching fractions,
while the latter affects the measurements of masses and widths of the resonances and the branching fractions of  the intermediate resonances.
The different sources of systematic uncertainty are discussed below.

We begin with the systematic uncertainty from event selection:
\begin{enumerate}[(i)]
\item The total number of $\psi(3686)$ events is obtained by studying inclusive
hadronic $\psi(3686)$ decays, giving a total uncertainty of 0.65\%~\cite{error:psipnum}.
This uncertainty is taken as a systematic uncertainty for this analysis.
\item The uncertainties of kaon tracking and PID efficiencies are estimated
using the control sample of $J/\psi \to K_{S}^{0}K^{*}$.
The difference between MC and data, $1.0\%$, is taken as the systematic uncertainty of the tracking or PID efficiency.
\item The $\overline{\Xi}^+$ reconstruction efficiency is studied with the control sample of
$J/\psi \to \Xi^{-}\overline{\Xi}^{+}$~\cite{error:xi}.
The difference between MC and data, $6.6\%$, is taken as the systematic uncertainty and
it includes the systematic uncertainties of MDC tracking and PID efficiencies for the $\bar{p}$ and both $\pi^+$.
\item In the fit to the RM$(K^-\overline\Xi^+)$ distribution,
 we consider three sources of uncertainty:
the signal model, the background model, and the fit range.
The signal model is changed to a double Gaussian function and
the background model is changed to a second-order polynomial function. In both cases, the change of the result is taken as  the systematic uncertainty.
The fit range is also changed to different values and the maximum change of the result is taken as  the uncertainty.
The total systematic uncertainty from fitting is the sum in quadrature of the three contributions.
\end{enumerate}

The second category is the systematic uncertainty from the PWA procedure:
\begin{enumerate}[(i)]

\item In this analysis,
the background is estimated from the $\overline \Xi^+$ sideband events in data.
We change the background level by $\pm 1\sigma$ and redo the PWA fit.
We also change the sideband range by $\pm 1\sigma$ and redo the PWA fit. For each variation, we take  the difference as the associated systematic uncertainty.
\item
We replace the non-res component in the nominal fit
by the other two processes (non-res$\rightarrow\Lambda\overline{\Xi}^{+}$
and non-res$\rightarrow{}K^{-}\overline{\Xi}^{+})$ and redo the PWA fit.
The differences are taken as the systematic uncertainty.
\item Besides $\Xi(1690)^-$ and $\Xi(1820)^-$ in the nominal fit, each known possible resonance has been included in the fit. Among them,  $\Xi(1620)^-$ is the most significant one with a statistical significance of 3$\sigma$. The difference between the fit results with and without $\Xi(1620)^-$ is taken as the systematic uncertainty  due to
possible additional resonances.
\item The systematic uncertainty associated with a change of the parameters $\lambda_{1}$ and $\lambda_{2}$ is evaluated by fixing $\lambda_{2}$ to 2.0 GeV and varying $\lambda_{1}$ between 1.5 GeV and 3.0 GeV. The maximum difference to the nominal result is taken as the systematic uncertainty.
\item  In analysis, the branching fractions are obtained with the optimized masses and widths of $\Xi(1690)$ and $\Xi(1820)$. Alternatively, the branching fractions are obtained with the masses and widths of $\Xi(1690)$ and $\Xi(1820)$ fixed to the PDG values.  The resulting changes in the measurements of branching fraction are assigned as systematic uncertainties.
\item The systematic uncertainty from fit bias is evaluated by applying nominal analysis procedure to signal MC samples generated according to the PWA results from data. The difference between input and output values is taken as the systematic uncertainty.
\end{enumerate}

The two categories of systematic uncertainties are listed in
Table~\ref{table:error_totalbr} and Table~\ref{table:error_totalxistars}.
\begin{table}[!htbp]
  \caption{Relative systematic uncertainties (in $\%$) on the branching fraction
  measurement of $\psi(3686)\rightarrow{}K^{-}\Lambda\overline{\Xi}^{+}+c.c.$.}
  \label{table:error_totalbr}
  \centering
  \begin{tabular}{cc}
    \hline
    \hline
    Source & Uncertainty (\%) \\
    \hline
    Number of $\psi(3686)$ events & 0.7 \\
    MDC tracking of $K^-$ & 1.0 \\
    PID of $K^-$ & 1.0 \\
    $\overline{\Xi}^{+}$ reconstruction & 6.6 \\
    Signal model & 0.8 \\
    Background shape & 0.1 \\
    Fit range & 0.3 \\
    \hline
    Total & 6.8 \\
    \hline
    \hline
  \end{tabular}
\end{table}
\begin{table*}[htbp]
  \caption{Systematic uncertainties on the measurements of the $\Xi^{*}$
  parameters and branching fractions.}
  \label{table:error_totalxistars}
  \centering
  \begin{tabular}{|c|c|c|c|c|c|c|}
    \hline
    \multirow{2}{*}{Source} & \multicolumn{3}{c}{$\Xi(1690)$} &
    \multicolumn{3}{|c|}{$\Xi(1820)$} \\ \cline{2-7}
    & $\Delta M $(MeV/$c^{2}$) & $\Delta \Gamma$ (MeV) & $\Delta {\mathcal B}/{\mathcal B}(\%)$
    & $\Delta M $(MeV/$c^{2}$) & $\Delta \Gamma$ (MeV) & $\Delta {\mathcal B}/{\mathcal B}(\%)$ \\
    \hline
    Number of $\psi(3686)$ events & - & - & 0.7 & - & - & 0.7 \\
    \hline
    MDC tracking of $K^{\pm}$ & - & - & 1 & - & - & 1 \\
    \hline
    PID of $K^{\pm}$ & - & - & 1 & - & - & 1 \\
    \hline
    $\overline{\Xi}^{+}$ reconstruction & - & - & 6.6 & - & - & 6.6 \\
    \hline
    Background level & 0 & 3 & 1.0 & 1 & 1 & 0.9 \\
    \hline
    Background sideband & 0 & 4 & 0.4 & 0 & 1 & 1.4 \\
    \hline
    Non-res component & 11 & 9 & 11.3 & 2 & 3 & 15.6 \\
    \hline
    Additional resonances & 5 & 17 & 14.0 & 2 & 2 & 3.7 \\
    \hline
    Different form factors & 0 & 2 & 3.0 & 1 & 8 & 1.3 \\
    \hline
    Fit bias & 1 & 3 & 8.4 & 0 & 1 & 2.4 \\
    \hline
    Resonance parameters of $\Xi(1690)$ and $\Xi(1820)$ & - & - & 19.8&- & -& 3.2\\
    \hline
    Total & 12 & 20 & 29.1 & 3 & 9 & 18.1 \\
    \hline
  \end{tabular}
\end{table*}

\section{SUMMARY AND DISCUSSION}
Based on $(448.1\pm2.9)\times10^{6}$ $\psi(3686)$ events collected with the
BES\uppercase\expandafter{\romannumeral3} detector at
BEPC\uppercase\expandafter{\romannumeral2} in 2009 and 2012,
we report the results of a partial wave analysis of $\psi(3686)\to{}K^{-}\Lambda\overline{\Xi}^{+}+c.c.$ Two excited hyperons,
 $\Xi(1690)^-$ and $\Xi(1820)^-$, are
observed in the $M(K^{-}\Lambda)$ and $M(K^{+}\overline{\Lambda})$ spectra.
Their masses, widths, spin-parities, and product branching fractions are measured.
The results obtained are summarized in Tables ~\ref{table:sum_reson} and ~\ref{table:sum_br}. 
We note that, whereas the masses of the $\Xi(1690)^-$ and $\Xi(1820)^-$ are in agreement with previous measurements, our width values are both larger than those measurements and only marginally consistent with them. However, our analysis is the first to use a PWA to include interference effects, and this might help explain the differences.
The spin-parities of $\Xi(1690)^-$ and $\Xi(1820)^-$ are measured for the first time, which are consistent with the quark model. This work improves the knowledge of the excited hyperon spectrum. To understand the internal structure of baryons and test theoretical predictions, further investigations with higher statistics are needed.

\begin{table}[H]
  \caption{Results obtained for $I(J^{P})$, mass and width for each component. The first (second) uncertainty is statistical (systematic).}
  \label{table:sum_reson}
  \centering
  \begin{tabular}{|c|c|c|c|}
    \hline
    Resonance & $I(J^{P})$ & M (MeV/$c^{2}$) & $\Gamma$ (MeV) \\
    \hline
    $\Xi(1690)^-$ & $1/2(1/2^{-})$ & $1685^{+3}_{-2}\pm{12}$ & $81^{+10}_{-9}\pm{20}$ \\
    $\Xi(1820)^-$ & $1/2(3/2^{-})$ & $1821^{+2}_{-3}\pm{3}$ & $73^{+6}_{-5}\pm{9}$ \\
    \hline
  \end{tabular}
\end{table}

\begin{table*}[htbp]
  \caption{Branching fraction results; the first (second) uncertainty is statistical (systematic).}
  \label{table:sum_br}
  \centering
  \begin{tabular}{|c|c|c|c|}
    \hline
    Resonance & Branching fraction \\
    \hline
    $\mathcal{B}(\psi(3686)\to{}\Xi(1690)^-\overline{\Xi}^{+} + c.c.)\times \mathcal{B}(\Xi(1690)^-\to
    {}K^{-}\Lambda)$ & $(1.06 \pm 0.10 \pm 0.31) \times 10^{-5}$ \\
    $\mathcal{B}(\psi(3686)\to{}\Xi(1820)^-\overline{\Xi}^{+} + c.c.)\times \mathcal{B}(\Xi(1820)^-\to
    {}K^{-}\Lambda)$ & $(1.78 \pm 0.10 \pm 0.32  ) \times 10^{-5}$ \\
    $\psi(3686)\to{}K^{-}\Lambda\overline{\Xi}^{+}+c.c.$ &
    $(3.60\pm0.10\pm0.24) \times 10^{-5}$ \\
        \hline
  \end{tabular}
\end{table*}

\textbf{Acknowledgement}

The BESIII Collaboration thanks the staff of BEPCII and the IHEP computing center for their strong support. This work is supported in part by National Key R\&D Program of China under Contracts Nos. 2020YFA0406300, 2020YFA0406400; National Natural Science Foundation of China (NSFC) under Contracts Nos. 11635010, 11735014, 11835012, 11935015, 11935016, 11935018, 11961141012, 12022510, 12025502, 12035009, 12035013, 12061131003, 12075250, 12075252, 12192260, 12192261, 12192262, 12192263, 12192264, 12192265; the Chinese Academy of Sciences (CAS) Large-Scale Scientific Facility Program; the CAS Center for Excellence in Particle Physics (CCEPP); Joint Large-Scale Scientific Facility Funds of the NSFC and CAS under Contract No. U1832207; CAS Key Research Program of Frontier Sciences under Contracts Nos. QYZDJ-SSW-SLH003, QYZDJ-SSW-SLH040; 100 Talents Program of CAS; The Institute of Nuclear and Particle Physics (INPAC) and Shanghai Key Laboratory for Particle Physics and Cosmology; ERC under Contract No. 758462; European Union's Horizon 2020 research and innovation programme under Marie Sklodowska-Curie grant agreement under Contract No. 894790; German Research Foundation DFG under Contracts Nos. 443159800, 455635585, Collaborative Research Center CRC 1044, FOR5327, GRK 2149; Istituto Nazionale di Fisica Nucleare, Italy; Ministry of Development of Turkey under Contract No. DPT2006K-120470; National Research Foundation of Korea under Contract No. NRF-2022R1A2C1092335; National Science and Technology fund; National Science Research and Innovation Fund (NSRF) via the Program Management Unit for Human Resources \& Institutional Development, Research and Innovation under Contract No. B16F640076; Polish National Science Centre under Contract No. 2019/35/O/ST2/02907; The Royal Society, UK under Contract No. DH160214; The Swedish Research Council; U. S. Department of Energy under Contract No. DE-FG02-05ER41374.

\end{document}